\documentclass[fleqn,usenatbib]{mnras}

\usepackage[T1]{fontenc}

\DeclareRobustCommand{\VAN}[3]{#2}
\let\VANthebibliography\thebibliography
\def\thebibliography{\DeclareRobustCommand{\VAN}[3]{##3}\VANthebibliography}

\usepackage{graphicx}	
\usepackage{amsmath}	
\usepackage{amssymb}	
\usepackage{caption}
\usepackage{subcaption}

\usepackage{newtxtext,newtxmath}

\title[Photosynthesis Under a Red Sun]{Photosynthesis Under a Red Sun: Predicting the absorption characteristics of an extraterrestrial light-harvesting antenna}

\author[C. D. P. Duffy et al.]{Christopher D. P. Duffy$^{1,2}$\thanks{E-mail: c.duffy@qmul.ac.uk}, Gregoire Canchon$^3$, Thomas J. Haworth$^3$\thanks{E-mail: t.haworth@qmul.ac.uk}, Edward Gillen$^3$, Samir Chitnavis$^{2}$ and 
\newauthor  
Conrad W. Mullineaux$^1$
\\
$^{1}$School of Biological and Behavioural Sciences, Queen Mary University of London, Mile End, London E1 4NS, UK\\
$^{2}$Digital and Environmental Research Institute, Queen Mary University of London, Empire House, Whitechapel, London E1 1HH, UK \\
$^{3}$School of Physical and Chemical Sciences, Queen Mary University of London, Mile End, London E1 4NS, UK\\
}

\date{Accepted XXX. Received YYY; in original form ZZZ}

\pubyear{2022}

\begin{document}
\label{firstpage}
\pagerange{\pageref{firstpage}--\pageref{lastpage}}
\maketitle

\begin{abstract}
Here we discuss the feasibility of photosynthesis on Earth-like rocky planets in close orbit around ultra-cool red dwarf stars. Stars of this type have very limited emission in the \textit{photosynthetically active} region of the spectrum ($400 - 700$ nm), suggesting that they may not be able to support oxygenic photosynthesis. However, photoautotrophs on Earth frequently exploit very dim environments with the aid of highly structured and extremely efficient antenna systems. Moreover, the anoxygenic photosynthetic bacteria, which do not need to oxidize water to source electrons, can exploit far red and near infrared light. Here we apply a simple model of a photosynthetic antenna to a range of model stellar spectra, ranging from ultra-cool (2300 K) to Sun-like (5800 K). We assume that a photosynthetic organism will evolve an antenna that maximizes the rate of energy input while also minimizing fluctuations. The latter is the \textit{noise cancelling} principle recently reported by Arp et al. 2020). Applied to the Solar spectrum this predicts optimal antenna configurations in agreement with the chlorophyll Soret absorption bands. Applied to cooler stars, the optimal antenna peaks become redder with decreasing stellar temperature, crossing to the typical wavelength ranges associated with anoxygenic photoautotrophs at $\sim 3300$ K. Lastly, we compare the relative input power delivered by antennae of equivalent size around different stars an find that the predicted variation is within the same order of magnitude. We conclude that low-mass stars do not automatically present light-limiting conditions for photosynthesis but they may select for anoxygenic organisms.     
\end{abstract}




\section{Introduction}
Over 5000 exoplanets (planets orbiting stars other than the Sun) have now been detected and confirmed \cite{ExoSurvey}. In addition to detecting planets, we are in some cases now also able to characterise their masses/radii and hence bulk composition \cite[e.g.][]{2007ApJ...669.1279S,2008ApJ...673.1160A, 2014ApJ...783L...6W} and through direct imaging and transmission spectroscopy probe their atmospheric compositions \cite[e.g.][]{2012ApJ...753..100B,2016ApJ...832..191N, 2022PASP..134i5003H}. With this information, we are getting ever closer to developing a census of possible habitats and it is becoming increasingly plausible that the presence of life on exoplanets might be inferred \citep{2002AsBio...2..153D, 2007AsBio...7...85S, 2017ARA&A..55..433K, doi:10.1089/ast.2017.1729}. 

Empirically, the distribution of stellar masses is well known to follow the stellar initial mass function \citep{2001MNRAS.322..231K, 2003PASP..115..763C}, wherein low mass stars are much more common. This, coupled to evidence \citep{2015ApJ...798..112M, 2019AJ....158..109H, 2020MNRAS.498.2249H} that terrestrial planetary occurrence rates are higher around low mass stars, implies that the majority of terrestrial planets orbit stars that are lower mass than the Sun ($T_{\textrm{eff}}<5780\,$K) \citep{Hsu2019,Hsu2020, Kristo2023_Terrestrial}. Though we note that the occurence rate of giants is lower around low mass stars \citep{2023arXiv230300659B} and the planet occurence rate is also sensitive to the stellar metallicity \citep[e.g.][]{2018AJ....156..221N, 2020AJ....160..253L}. 

\begin{figure*}[ht]
\centering
\includegraphics[width=\linewidth]{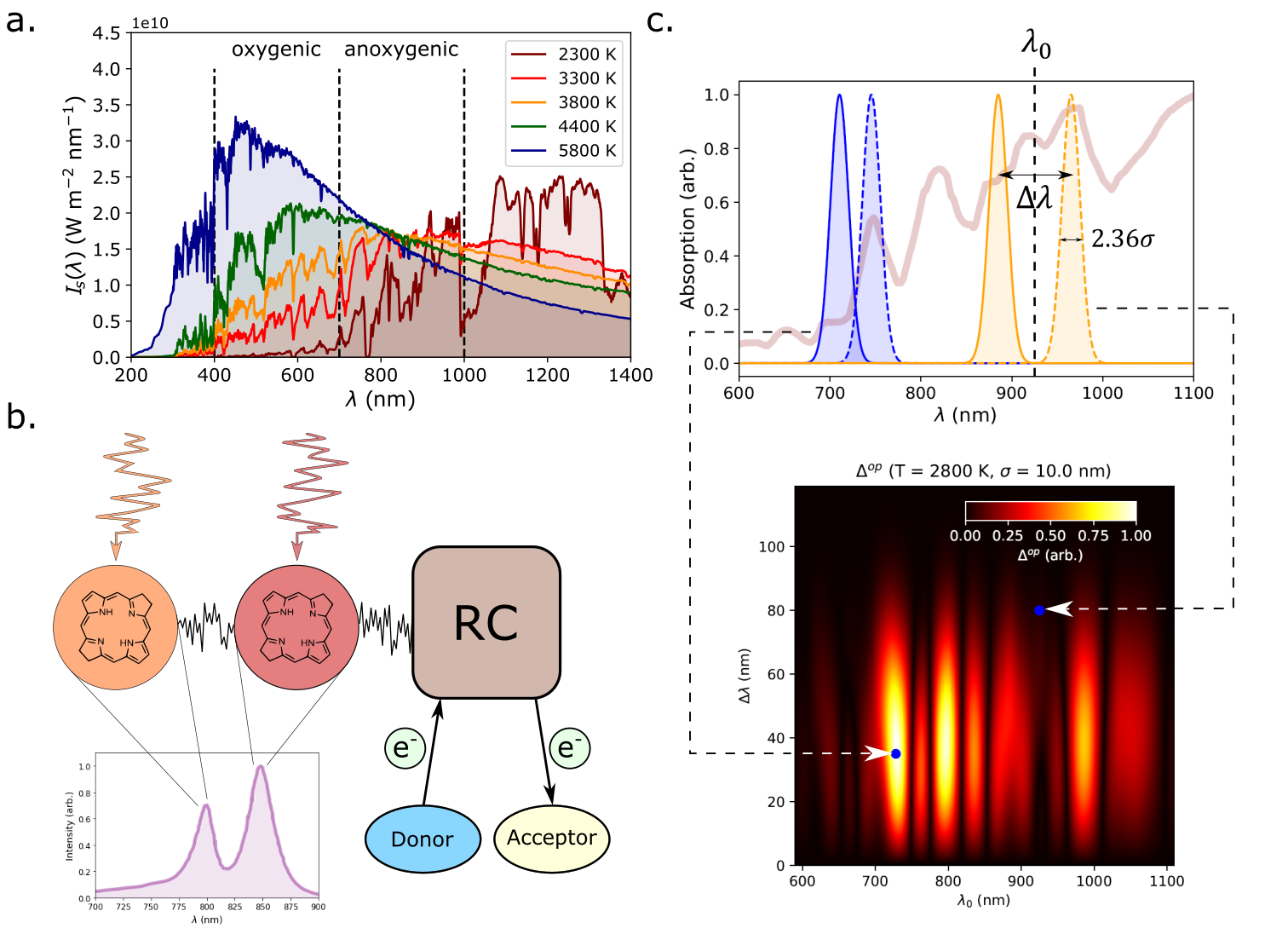}
\caption{\textbf{a.} Model surface spectral flux, $I_{s}(\lambda;T_{s})$ , for planets within the middle of the habitable zone around parent stars of different effective temperatures, $T_{s}$. The spectral fluxes are taken from PHOENIX radiative transfer models \citep{2013A&A...553A...6H} of surface spectral flux and the habitable distances are estimated according to a simple radiative equilibrium model outlined in the Methodology. A sparse range of temperatures is shown purely for clarity and the vertical dashed lines approximately demarcate the absorption regions for oxygenic and anoxygenic photosynthesis, with the former often referred to as \textit{photosynthetically active radiation} (PAR). \textbf{b.} A schematic diagram of the concept of a dual-input noise-cancelling antenna. Two sub-populations of pigments with similar (but different) absorption maxima funnel energy to the reaction centre (RC) which oxidizes an electron donor and reduces an acceptor. The two absorbing populations tend to operate in series (e.g. Chl \textit{b} transferring energy to Chl \textit{a} in plant antenna complexes) and are subject to both external and internal noise. The former reflects the highly dynamic nature of the light-environment while the latter results from fluctuations of the energy transfer pathways within the antenna. \textbf{c.} An example of the matrix representation of $\Delta^{op}\left(\lambda_0, \Delta\lambda\right)$ for a fixed value of absorber width, $\sigma=10$ nm. Above this is an illustration of two examples of antenna configuration superimposed on the $2800$ K spectrum (dark red). $T=2800$ K is chosen here purely for illustrative purposes as it exhibits many sharp bands of optimal (and extremely sub-optimal) antenna configurations. Note the distinction between the standard deviation or 'width', $\sigma$, and the Full Width at Half Maximum, $\Gamma~2.63 \sigma$ of the Gaussian peak. }
\label{fig:Intro_figure}
\end{figure*}

A number of terrestrial planets have been discovered around low mass stars on compact orbits, including well-known examples such as Trappist-1 \citep{2017Natur.542..456G},  Proxima Centauri  \citep{2016Natur.536..437A, 2022A&A...658A.115F} and LHS 1140 \citep{2017Natur.544..333D}.  This includes ``habitable zone'' planets where it is expected that water would be in the liquid phase on the surface of a rocky planet \citep{1993Icar..101..108K, 2018A&A...613A..68G, 2017Natur.544..333D}.  Trappist-1 in particular has multiple planets in the habitable zone, making it is a high priority target with over 200 hours of observing time allocated to that planetary system in JWST cycle 1. However to understand habitability we need to go beyond the simple definition of supporting liquid water.
***
Currently, Earth's biosphere is largely supported by photosynthesis, with oxygenic photoautotrophs specifically being responsible for $>99\%$ of the global primary production \citep{Overmann2013}. This does, however, depend on the biome, with anoxygenic photosynthesis contributing $\sim 30\%$ in ancient, sulphide-rich lakes \citep{Overmann2013}, and chemosynthetic microbes becoming the dominant producers in desert ecosystems \citep{Bay_chemo}. Previously, it was argued that evolution of multi-cellular life, particularly animals, was only possible in an an atmosphere oxygenated by photosynthesis, though this view is currently being challenged \citep{Cole_Oxygen,Bozag_Oxygen}. Regardless, oxygenic photosynthesis currently covers $50-80\%$ of the Earth's surface \citep{Field_Photosynth_1998}, in the form of canopy forests, savannah, grasslands and marine cyanobacteria and algae, and is detectable, at least at interplanetary distances, as a sharp increase in solar reflectance for wavelengths $\lambda > 700-750$ nm \citep{Sagan1993VRE,Arnold_VRE2002}. This `vegetation red edge' (VRE) is therefore a promising potential biosignature to look for in exo-planet surveys \citep{Seager2005VRE}, though it may be hidden in the case of life beneath substrates \citep{2009AsBio...9..623C}. Biochemically it arises from chlorophyll \textit{a}, the pigment responsible for water oxidation and quinone reduction in Photosystem II (PSII), and NADP reduction in Photosystem I (PSI), photochemical processes that require quanta of $\lambda = 680$ and $700$ nm respectively. While higher energy photons ($400 -700$ nm) are readily utilized, since excess energy is shed non-radiatively in the antenna \citep{vanAmerongenLHCII}, redder photons are insufficiently energetic to drive charge separation. This would imply that low mass stars, with limited emission in the `photosynthetically active' region of $400 < \lambda < 700$ nm (see Fig. \ref{fig:Intro_figure} \textbf{a.}), are unlikely to the support complex biospheres \citep{CovoneOxygenic2021}. However, this definition of `photosynthesis' is perhaps too narrow. Several species of cyanobacteria bind red-shifted Chl \textit{d} and \textit{f}, with absorption maxima at $790 - 800$ nm, as an adaptation to light that is strongly attenuated in the $400 - 700$ nm region by the environment and other organisms \citep{ViolaRedPhoto2022,Tros2021RedLimit}. Moreover, anoxygenic photoautotrophs, such as purple \citep{hu_ritz_damjanović_autenrieth_schulten_2002} and green sulphur bacteria \citep{Gregersen_Green_Sulfur} utilize light in the $800 - 1000$ nm region. The fact that they source electrons from more readily oxidizable compounds, such as hydrogen sulphide, ferrous iron, hydrogen, etc., relaxes the requirement for quanta of $\lambda < 700-800$ nm \citep{Bryant2006Prokaryote}, though does limit their habitat. It was previously assumed that anoxygenic photosynthesis evolved first, with oxygenic photo-autotrophs evolving around the Great Oxidation Event \citep{LyonsGOE}, although this has been challenged recently, with compelling genetic \citep{Cardona2019} and geological \citep{Wang2018Oxygenic} evidence for a much earlier appearance. Even so, recent modelling implies that Earth's VRE signal may have risen to its current level as recently as only 500 Mya, which may limit the likelihood of VRE detection to older, warmer Earth-like planets \citep{doi:10.1089/ast.2017.1798}. There is therefore significant motivation to broaden our search to redder, more exotic forms of photosynthesis.    

It is assumed that the evolution of photosynthetic processes and structures was largely dictated by the intensity, spectral quality and dynamics of the available light. An early adaptation was the antenna-reaction centre architecture, which massively enhances the absorption cross-section of the photosystems and allows for dynamic regulation of energy input in a fluctuating light environment \citep{WolfeAntenna1994,FlemingPrinciples2012}. Physically, the antenna is a massive, modular assembly of light-harvesting pigment-protein complexes. Their function is to capture light and transfer the resulting excited electronic state or `exciton' to the reactions centres (which they do with near-total quantum efficiency). Antenna proteins generally bind several types of pigments in order to provide broad spectral coverage, creating a `funnel' structure in which higher energy pigments donate to lower energy ones. For example, the major light-harvesting complex of PSII in higher plants, LHCII, binds Chl \textit{a} and \textit{b} along with several carotenoids \citep{Liu_LHCII_PSII}. Nevertheless, antenna complexes are not black bodies, instead possessing well-defined and often narrow absorption peaks. 

\cite{Kiang2007} published a comprehensive review of the relationship between the absorption profiles of Earth's photo-autotrophs and their respective local light environments. Based on empirical trends they proposed several rules: (1) The absorption peak of the antenna is located close to the local irradiance maximum, to maximize energy input. (2) The absorption peak of the reaction centres are close to longest wavelength in the irradiance range, since excitons must be funneled from a higher energy antenna. (3) Accessory pigments (such as Chl \textit{b} or carotenoids in plants) will absorb towards the shortest wavelength of the irradiance window, since they must funnel excitons to the primary pigments (e.g. Chl \textit{a}) in the antenna. 

Following on from \cite{Bjorn1976}, \cite{Marosvolgyi2010} applied a different approach, considering the balance between the need to absorb as much light as possible with the potentially prohibitive metabolic cost of synthesizing and maintaining a vast array of different pigment co-factors. The model depends on a \textit{cost parameter}, $C$, which reflects the fraction of captured energy that is used to synthesize, regenerate and repair the light-harvesting antenna. Choosing $C \approx 0.96$ reproduces the red absorption band of boht the chloroplast and the chromatophore of purple bacteria in their respective native light environments. The argument is that there is an optimum band in which to harvesting photon. Harvesting bluer photons carries the burden of synthesizing additional pigments, while harvesting redder photons is less efficient due to an increased likelihood of ambient thermal excitation of the the antenna pigments (making stimulated emission rather than absorption more likely). This seems to miss the distinct blue absorption band of the chloroplast which is composed of the Chl \textit{a} and \textit{b} Soret bands and the $S_{2}$ bands of various carotenoids. They argue that Chl \textit{a} and \textit{b} were selected for their red $Q_{y}$ bands, with their Soret bands being a (fortunate) side effect of their electronic structures. Similarly, they argue that carotenoids were primarily selected for their structural, antioxidant and triplet-quenching properties. One criticism of this model is that the predicted spectrum is extremely sensitive to the choice of $C$. For $C=0.96$ the predicted antenna spectrum is in very good agreement with the chloroplast $Q_{y} $ band, even down to the stoichiometry between Chl \textit{a} and \textit{b}. However, $C=0.75$ yields a single absorption band covering the entire $500 - 750$ nm region of the spectrum, making it almost black. At $C=0.5$ it absorbs more-or-less uniformly across $400 - 900$ nm. Moreover, $C=0.96$ seems to imply that nearly all of the light energy captured and converted by the organism is re-invested in synthesizing antenna pigments, though this parameter should not be over-interpreted and the fact that the same value of $C=0.96$ was found for both plants and purple bacteria indicates this model captures some fundamental feature of the antenna.   

More recently, \cite{Arp2020} took a different approach, arguing that the absorption profile of the antenna was primarily determined by a requirement for \emph{noise cancellation}. An antenna is subject to noise, both externally as fluctuations in spectral intensity, and internally due to the stochastic nature of the branching exciton transfer pathways. This risks periods where the photosystems are either under-powered, which reduces growth rate, or over-powered, which risks photodamage. \cite{Arp2020} argue (very convincingly) that this noise is minimized by a two-channel antenna, composed of a pair of Gaussian absorbers with similar (but different) absorption maxima, located where the gradient of the spectrum is steepest \citep[see Fig. \ref{fig:Intro_figure} \textbf{b.,}][]{Arp2020}. This principle seems to predict the $Q_{y}$ and Soret profiles of plants, purple bacteria and green sulfur bacteria, based solely on the the spectrum of locally available light, though it should be noted that these solutions are neither unique nor necessarily optimal (see below). 


These different approaches are complementary rather than contradictory, and the relationship between antenna structure and the local light environment is likely dependent on multiple factors. Since this relationship is based (largely) on fundamental rules of photo/redox chemistry, it is reasonable to assume it is universal. 

In 2002 Wolstencroft and Raven modelled the rate of oxygenic photosynthesis on the surface of habitable-zone planets, with and without cloud cover, orbiting a range of stars \citep{Wolstencroft2002}. Assuming a model absorption profile taken from the model species \textit{Nerium oliander}, they found that oxygenic photosynthesis performed best on cloudless planets orbiting F-type stars ($6000-7500$ K) and very poorly around cooler stars (K and M-type). The reason for the latter was a mismatch between the spectral irradiance and the \textit{N. oliander} absorption profile. They then proposed that oxygenic photosynthesis around these low mass stars may still be feasible via multi-photon processes, in which two or more $\lambda >1000$ nm photons deliver the required energy as opposed to a single $\lambda<700 nm$ photon. \cite{Tinetti2006DetectabilityOR} explored this idea further and proposed that such processes could produce a red-shifted VRE in the Near Infrared (NIR, $750<\lambda<1500$ nm).  

In a companion paper to the one discussed above \cite{Kiang2007_Part2} applied their empirical antenna-irradiance relationship to model irradiances for Earth-like planets orbiting in the habitable zones around F, K and M-type stars. They predicted that F2V stars ($\sim 1.5 \text{M}_{\odot}$) would favour photosynthetic pigments that absorb in the blue, K2V stars ($\sim 0.8 \text{M}_{\odot}$) would favour red-orange (as they do on Earth), and M-type stars (red dwarf stars, $0.08-0.6 \text{M}_{\odot}$) would favour absorbers in several NIR bands ($930-1100$ nm, $1100–1400$ nm, $1500–1800$ nm, and $1800–2500$ nm). The reason for the multiple bands is that irradiance in the NIR is strongly modified by atmospheric transmission, leading to multiple, distinct bands in the local spectral flux \citep{Segura2003}. 

More recently, \cite{LehmerFarRedEdge2021} applied the model of \cite{Marosvolgyi2010} discussed above to similar model irradiances. Assuming that the $C=0.96$ cost parameter that applies to Earth organisms is universal, they predicted similar antenna absorption profiles to \cite{Kiang2007_Part2}. The one difference was that even for very low mass M5V stars ($\sim 0.16 \text{M}_{\odot}$) the model did not predict absorption in NIR bands at $1100–1400$ nm, $1500–1800$ nm, and $1800–2500$ nm, instead predicting a broad absorption peak at $1050$ nm with a almost negligible shoulder predicted at $987$ nm. By the logic of the \cite{Marosvolgyi2010} model, background thermal excitation of the antenna pigments would become significant at these longer wavelengths, making light-harvesting less efficienct.  

Finally, at time of writing, \cite{CHall_Habitable_Preprint} have published a model in which they define the \textit{photosynthetic habitable zone} as the range of orbital radii in which both liquid water can occur and oxygenic photosynthesis is feasible. The model assumes that the maximal rate of photosynthesis, $P_{\text{max}}$, is dependent on surface irradiance (via a set of empirical parameters derived from phytoplankton \citep{Yang2020Phytoplankton}), ambient surface temperature, and the dark respiration rate, $R_{\text{rate}}$. $R_{\text{rate}}$ is essentially the steady-state energy requirement of the organism merely to continue existing, including the metabolic burdens of protein turnover, repair, maintaining homeostasis, etc. To enable growth, reproduction, etc. photosynthesis must generate a surplus to $R_{\text{rate}}$. On Earth $R_{\text{rate}} \leq 0.3P_{\text{max}}$, meaning that less than a third of the photosynthetic yield needs to be reinvested in base-level survival, reflecting generally favourable conditions for life on Earth \citep{GeiderR_rate}. If conditions are similarly favourable on other Earth-like planets then oxygenic photosynthesis may be feasible around stars of $>0.6 \text{M}_{\odot}$. This range can be extended to even smaller stars in the absence of any atmospheric attenuation of light or any greenhouse effect (though the authors admit this may be extremely optimistic).  

As with \cite{Arp2020} it may be instructive to apply a different set of criteria to this problem. Here we apply a modified form of their noise cancelling antenna principle to model spectral fluxes of a range of stars, from very low mass/temperature M-dwarfs to something more like the Sun. We hypothesize that the antenna absorption profile will evolve to minimize input noise while maximizing the total input power.   

\section{Methodology}
\medskip
\subsection{Model Stellar/Solar Spectra}
\medskip
We use stellar spectral models for stars of different masses/temperatures generated by the \textsc{phoenix} radiative transfer code \citep{2013A&A...553A...6H}. These are typically over one and half million lines in length, so we smooth and re-sample the spectrum down to 4000 points, which still captures the large scale features and reduces the computation time for calculating the optimal antenna absorption characteristics (see Fig. \ref{fig:Intro_figure} \textbf{a.}). Smoothing features over wavelength ranges smaller than the smallest absorption band we consider is justified, given that all radiation across the band is absorbed. In other words, antenna bands cannot resolve spectral variations on scales smaller than their own effective width.

The standard solar spectrum used in Fig. \ref{fig:Solar_peaks} is an AM1.5 spectrum generated by the latest version of the SMARTS (Simple Model of the Atmospheric Radiative Transfer of Sunshine) code \citep{Gueymard2004SMARTS} and published by NREL \citep{Smarts}

\subsection{Determining optimal antenna absorption characteristics for a given stellar spectral flux}
\medskip
Here we describe how we construct our basic model as a noise cancelling system that maximises the total power input, as discussed above. In line with the previous work of \cite{Arp2020} we assume that the antenna absorption profile is comprised of a closely spaced Gaussian doublet (see Fig. \ref{fig:Intro_figure} \textbf{b.} and \textbf{c.}),
\begin{equation}\label{eq:1}
    A\left(\lambda;\lambda_{0},\Delta\lambda,\sigma\right)={\frac{1}{2}}\left(A_{+}\left(\lambda;\lambda_{0},\Delta\lambda,\sigma\right)+A_{-}\left(\lambda;\lambda_{0},\Delta\lambda,\sigma\right)\right)
\end{equation}
where,
\begin{equation}\label{eq:2}
    A_{\pm}\left(\lambda;\lambda_{0},\Delta\lambda,\sigma\right)=\frac{1}{2\sigma\sqrt{2\pi}}\exp\left(-\frac{\left(\lambda-\lambda_{0}\pm\frac{\Delta\lambda}{2}\right)^{2}}{2\sigma^{2}}\right)
\end{equation}
$\lambda_{0}$ is the central wavelength of the absorber pair, $\Delta\lambda$ is the separation between peaks, and $\sigma$ is the standard deviation of the Gaussian curve, hereafter the `width' (see Fig. \ref{fig:Intro_figure} \textbf{c.}). To limit the number of free parameters we assume that $A_{+}$ and $A_{-}$ have the same width and amplitude. $A_{+}$ and $A_{-}$ represent the two input channels of the antenna and we define the input power of each channel as,
\begin{equation}\label{eq:3}
    P_{\pm}=\int_{0}^{\infty} d\lambda \sigma_{0}\frac{hc}{\lambda}A_{\pm}\left(\lambda;\lambda_{0},\Delta\lambda,\sigma\right)I_{s}\left(\lambda;T_{s}\right)
\end{equation}
where $\sigma_{0}$ is the integrated optical cross-section of the antenna, $I_{s}\left(\lambda\right)$ is the incident spectral flux (see next sub-section), $h$ is the Planck constant and $c$ is the speed of light. \cite{Arp2020} showed that the requirement for a noise-cancelling antenna is satisfied by maximizing the \textit{power input difference},
\begin{equation}\label{eq:4}
    \Delta^{op}\left(\lambda_{0},\Delta\lambda,\sigma\right)=\frac{1}{2}\left|P_{+}-P_{-}\right|
\end{equation}
subject to the constraint that $\Delta\lambda<6\sigma$. They provide a thorough mathematical justification of this and the interested reader is directed to the original article \citep{Arp2020}. Here we assume that there will be an additional selection pressure to maximize the total power input of the antenna,
\begin{equation}\label{eq:5}
    P_{in}\left(\lambda_{0},\Delta\lambda,\sigma\right)=\frac{1}{2}\left(P_{+}+P_{-}\right)
\end{equation}
particularly in the dim, red light environment provided by M dwarf stars. We calculate the matrices $\Delta^{op}\left(\lambda,\Delta\lambda\right)$ and $P_{in}\left(\lambda,\Delta\lambda\right)$ for widths $\sigma=10.0,15.0,20.0,25.0$,and $30.0$ nm, which is reasonably representative of the range of widths observed for photosynthetic pigments on Earth. For example, the $0-0$ peak of the $Q_y$ band of Chl \textit{a} in solution is approximately Gaussian with width $\sigma \approx 12$ nm \citep{KnoxSpring2003}, while the vibronic peaks of carotenoids such as lutein can have $\sigma \approx 20-40$ nm \citep{Gray2020LHCII}. $\Delta^{op}$ and $P_{in}$ are then re-scaled so that their maximal values across the entire parameter space equal unity, allowing us to define a product function,
\begin{equation}\label{eq:6}
\begin{aligned}
    \Pi_{in}\left(\lambda_{0},\Delta\lambda,\sigma\right)&=\tilde{\Delta}^{op}\left(\lambda_{0},\Delta\lambda,\sigma\right)\tilde{P}_{in}\left(\lambda_{0},\Delta\lambda,\sigma\right)\\
    &=\left|\left(\tilde{P}_{+}\right)^{2}-\left(\tilde{P}_{-}\right)^{2}\right|
\end{aligned}
\end{equation}
where the tilde indicate re-scaled functions. The definition of $\Pi_{in}$ means that an antenna configuration with $\Pi_{in}~1.0$ satisfies both the noise cancellation and power input maximization conditions, while $\Pi_{in}~0.0$ means one (or both) of these conditions is not well satisfied.   

We therefore generate a set of matrices or maps (each corresponding to a different width, $\sigma$) for $\tilde{\Delta}^{op}$, $\tilde{P}_{in}$ and $\Pi_{in}$ (see Fig. \ref{fig:5800K_maps_peaks}). These are then analysed to identify the most topologically significant peaks via the method of persistence homology \citep{Otter_Homology_2017,Taskesenfindpeaks2020}. For this particular problem - identifying local maxima on a 2D surface - this has a rather intuitive interpretation. Taking one of the maps shown in Fig. \ref{fig:5800K_maps_peaks} we can scan the z-axis (in our case $\Delta^{op}$, $P_{in}$, or $\Pi_{in}$) from $1 \rightarrow 0$. As we do so peaks will appear, initially as distinct 'islands', before merging together. The \textit{persistence score} of a peak is then the difference between the z-values at which it appears and when it merges with other peaks. A very tall, very sharp peak, well-separated from its neighbours, will have a high persistence score, while short, broad, clustered peaks will merge with each other very soon after they appear.

\subsection{Comparing power input for optimal antenna configurations around different stars}
\medskip
Having calculated the optimal antenna configurations for given stellar temperature, we then compare the input powers across different temperatures. We define the ratio,

\begin{multline}\label{eq:7}
    \phi\left(\lambda_{0},\Delta\lambda,\sigma; T\phantom{x} |\phantom{x} \lambda_{0}' ,\Delta\lambda',\sigma';T' \right)= \\ \frac{P_{in}^{T}\left(\lambda_{0},\Delta\lambda,\sigma\right)}{{\sigma_{0}^{T }}}\cdot\frac{\sigma_{0}^{T'}}{P_{in}^{T'}\left(\lambda_{0}' ,\Delta\lambda',\sigma'\right)}
\end{multline}
where $T$ and $T'$ denote specific stellar temperatures. We will assume $T'=5800 K$, effectively treating the Earth-like antennae as our reference configurations. To constrain the parameters space we need to explore we will assume that the total cross-sections are equivalent, $\sigma_{0}^{T}=\sigma_{0}^{T'}$ but we will discuss the implications of this assumption below. 

Eqn. (\ref{eq:7}) depends on \textit{relative} spectral irradiance of two antennae evolving around different parent stars. Formally, $I_s(lambda;T)$ are spectral flux densities  W m$^{-2}$ nm$^{-1}$ at the stellar surfaces, so we therefore estimate the spectral flux at the planetary surface by assuming approximate habitable distance for each start type. Assuming that both the star and the orbiting planet are spherical black bodies, then the condition of radiative equilibrium implies the relation,

\begin{equation}\label{eq:8}
\left(\frac{R_{s}}{A_{p}}\right)^{2}=4\left(\frac{T_{p}}{T_{s}}\right)^{2}
\end{equation}
where $T_{\star}$ and $T_{p}$ are the equilibrium temperatures of the star and planet respectively, $R_{s}$ is the stellar radius, and $A_{p}$ is the planetary orbital radius. We then estimate $A_{p}\left(T_{p};R_{s},T_{s}\right)$ for a habitable (liquid water) temperature range $273 \text{K} \leq T_{p}\leq 373 \text{K}$. We then estimate the planetary surface spectral flux as,

\begin{equation}\label{eq:9}
I_{s}\left(\lambda;\langle A_{p}\rangle, R_{s},T_{s}\right)=F_{s}\left(\lambda;R_{s},T_{s}\right)\left(\frac{R_{s}}{A_{p}}\right)^{2}
\end{equation}
where $F_{s}\left(\lambda;R_{s},T_{s}\right)$ are the stellar surface spectral fluxes given by the PHOENIX radiative transfer models. A representative selection of $I_{s}(\lambda)$ are shown in Fig. \ref{fig:Intro_figure} \textbf{a.} 

\section{Results}

\subsection{Noise reduction and power maximization for the Solar spectrum}

\begin{figure*}
\centering
\includegraphics[width=0.8\linewidth]{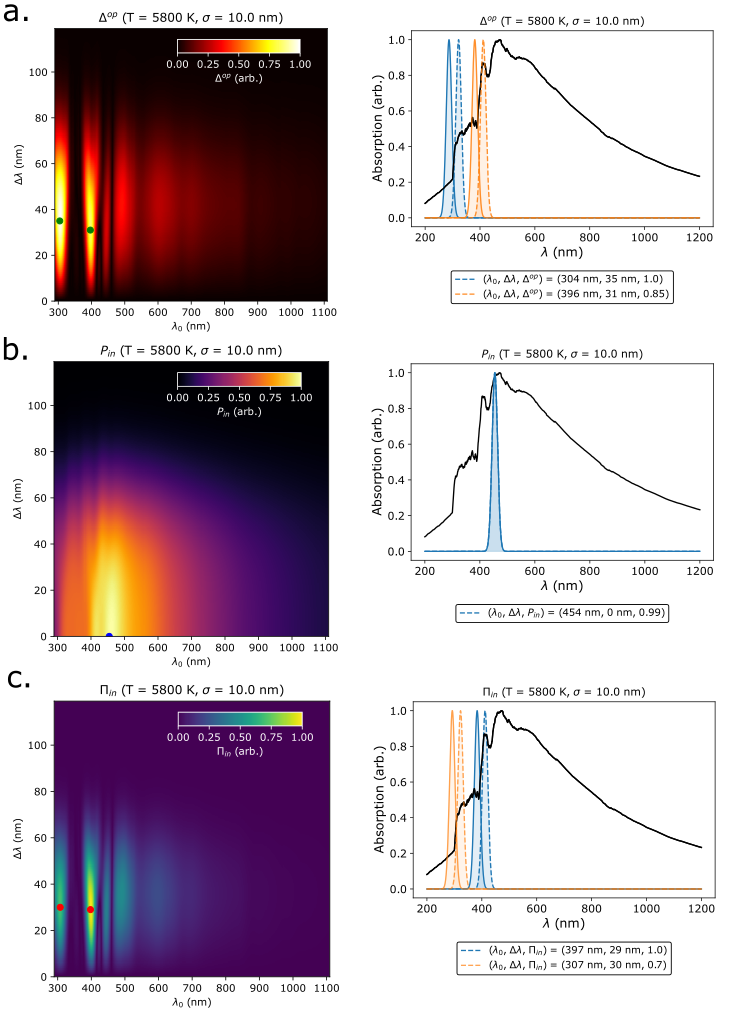}
\caption{Optimal antenna properties for a $T_{s} = 5800$ K PHOENIX spectral flux. The noise cancelling conditions requires an antenna absorption profile that is a Gaussian doublet with central wavelength $\lambda_0$ and separation $\Delta \lambda$. A width of $\sigma = 10 nm$ gives the optimal values but a range of widths are shown in the Supplementary Material. \textbf{a. (left)} A heatmap of the noise cancelling parameter, $\Delta^{op}\left(\lambda_{0},\Delta\lambda,w\right)$, as a function of $\lambda_0$ and $\Delta \lambda$. The green dots indicate the most topologically distinct peaks as determined by the persistence score. \textbf{a. (right)} The optimal peaks shown with the spectrum (black). \textbf{b.} As a. but for the total power input, $P_{in}\left(\lambda_{0},\Delta\lambda,w\right)$. The most topologically distinct peak is indicated with a blue dot but the dependency of $P_{in}$ on $\lambda_{0}$ and particularly $\Delta\lambda$ are much flatter then for $\Delta^{op}$. \textbf{c.} As a. but for the product function $\Pi_{in}\left(\lambda_{0},\Delta\lambda,w\right)$. For $\Delta^{op}$ and $\Pi_{in}$ we note that there are several smaller-amplitude, less-distinct peaks on the red edge of the spectral flux}.
\label{fig:5800K_maps_peaks}
\end{figure*}

To validate the model of \cite{Arp2020} we first apply our model to the $T_{s}=5800$ K spectral flux which is representative of the Solar spectrum at the top of Earth's atmosphere (Fig. \ref{fig:5800K_maps_peaks}). Fig. \ref{fig:5800K_maps_peaks} \textbf{a.} (left) shows the noise cancelling parameter $\Delta^{op}\left(\lambda_{0},\Delta\lambda\right)$ for absorber widths $\sigma = 10$ nm. The latter was chosen since it gave the absolute optimal solution but the same data for larger $\sigma$ are listed in the Fig. S1 Supplementary Material. We see that $\Delta^{op}$ is more strongly-dependent on $\lambda_{0}$ than on $\Delta\lambda$. As expected from \cite{Arp2020}, the values of $\left(\lambda_{0},\Delta\lambda\right)$ that maximize $\Delta^{op}$ are those that span the regions of $I_{s}(\lambda)$ with the steepest gradient, with two optimal solutions on the blue edge of $I_{s}(\lambda)$, centred on $\lambda_{0}=304$ and $396$ nm (see Fig. \ref{fig:5800K_maps_peaks} \textbf{a} (right)).  

Fig. \ref{fig:2800K_maps_peaks} \textbf{a.} (left) shows $P_{in}\left(\lambda_{0},\Delta\lambda\right)$ for $\sigma = 10$ nm, revealing less of a dependence on $\lambda_{0}$ and $\Delta\lambda$ than $\Delta^{op}$.  The requirement to maximize $P_{in}$ results in a single absorption line ($\Delta\lambda=0$ nm) located at the spectral maximum. One would expect maximizing $P_{in}$ would select for large $\sigma$ (see Fig. S2 of the Supplementary Material), to cover as much of the incident spectral flux as possible, but the normalization condition in Eqn. (\ref{eq:2}) means that broader absorption lines have smaller peak amplitudes. While this may seem like an artificial model constraint it is essentially an assumption that increasing $\sigma$ is predominantly via increasing inhomogeneous broadening, i.e. increasing the variance in the absorption peaks of chemically-identical pigments (due to an anisotropic solvent protein environment) rather than increasing the line width of the pigment itself (homogenous broadening). 

Fig. \ref{fig:5800K_maps_peaks} \textbf{c.} (left) shows $\Pi_{in}\left(\lambda_{0},\Delta\lambda\right)$ for $\sigma = 10$ nm, which is qualitatively similar to $\Delta^{op}\left(\lambda_{0},\Delta\lambda\right)$. The key difference is that simultaneously maximizing both $P_{in}$ and $\Delta^{op}$ favours the antenna configuration closer to the peak of $I_{s}$ (see Fig. \ref{fig:5800K_maps_peaks} \textbf{c.} (right)). $\Pi_{in}\left(\lambda_{0},\Delta\lambda\right)$ for larger $\sigma$ are shown in the Fig. S3 of the Supplementary Material.  

\begin{figure*}
\centering
\includegraphics[width=0.5\linewidth]{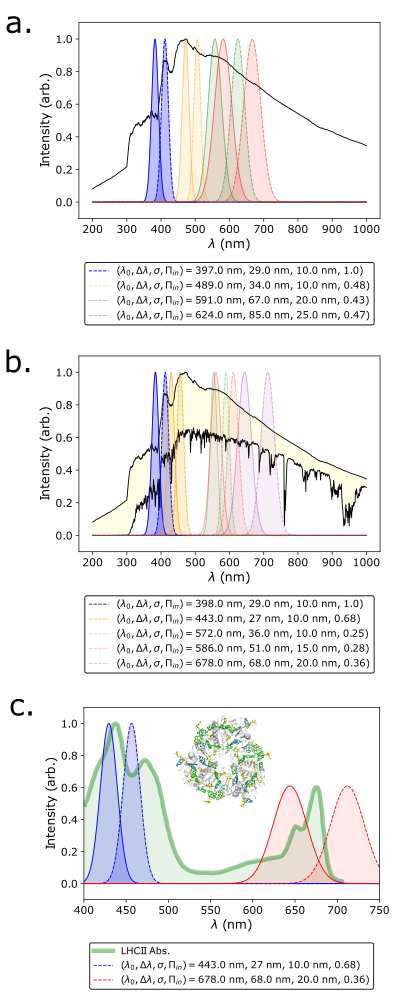}
\caption{\textbf{a.} Absorption doublets corresponding to local peaks in $\Pi_{in}$ on the red side of a $T_{s}=5800$ K PHOENIX spectral flux, $I_{s}(\lambda)$. These peaks are less topologically distinct than the absolute optimum (shown in blue) and less effective at maximizing noise cancellation $\Delta^{op}$ and power input, $P_{in}$. \textbf{b.} Same as a. but for a standard Solar spectrum with AM1.5 atmospheric filtering. As well as adding another peak (orange) on the blue side of the spectrum, the overall flattening in the $\lambda > 500 nm$ region reduces the $\Pi_{in}$ scores for doublets on the red side. \textbf{c.} Two absorption doublets from b. shown against the absorption spectrum of major plant light-harvesting complex LHCII (green, structure of LHCII shown inset).}
\label{fig:Solar_peaks}
\end{figure*}

Fig. \ref{fig:5800K_maps_peaks} suggests that the optimal light-harvesting system for the Solar spectral flux would be composed of a pair of $\sigma=10$ nm absorbers centred on $385$ and $415$ nm respectively. This is qualitatively similar to the Chl \textit{a} and \textit{b} blue absorption (or 'Soret') bands at approximately  $450$ and $425$ nm respectively but are not an exact match. 
Crucially, this model seems to miss the $680$ and $650$ nm red (or '$Q_y$') bands of Chl \textit{a} and \textit{b} respectively, though we note there are several smaller peaks in $\Delta^{op}$ and $\Pi_{in}$ on the red side of $I_{s}$ (Fig. \ref{fig:5800K_maps_peaks} \textbf{a.} and \textbf{c.}). Three such configurations, with $\Pi_{in} \sim 0.5$, are show alongside the absolute optimal configuration in Fig. \ref{fig:Solar_peaks} \textbf{a.}. These less optimal solutions cover the range $490 \text{nm} < \lambda_{0} < 625 \text{nm}$. 

For Earth-based photosynthesis we have the benefit of very accurate atmospheric transmission data. Fig. \ref{fig:Solar_peaks} \textbf{b.} shows how the position and height of the $\Pi_{in}$ peaks change for an AM1.5 standard solar spectrum. The absolute optimum peak is largely unaffected but the peak at $\lambda=489$ nm in Fig. \ref{fig:Solar_peaks} \textbf{a.} is blue-shifted to $\lambda_{0}=443$ nm. The peaks on the red side of the spectrum move a little and their relative $\Pi_{in}$  scores decrease due to the overall gradient of the spectrum, and therefore $\Delta^{op}$, decreasing. The $\tilde{\Delta}^{op}$, $\tilde{P}_{in}$, and $\Pi_{in}$ matrices for the full range of widths are included in Figs S4-S6 of the Supplementary Material, which show that the number and density of distinct bands in $\Delta^{op}$ and $\Pi_{in}$ increases dramatically relative to the $5800$ K spectrum. This simply reflects the fact that the AM1.5 spectrum is contains a number of sharp features on its red edge (see Fig. \ref{fig:Solar_peaks} \textbf{b.}).

Finally, Fig. \ref{fig:Solar_peaks} \textbf{c.} shows the absorption spectrum of LHCII, the major light-harvesting protein of vascular plants (spectrum digitized from \cite{LHCII_spectrum}, structure shown inset), with two of the antenna configurations identified in Fig. \ref{fig:Solar_peaks} \textbf{b.} The broad peaks in the region $400 - 500$ nm of the LHCII absorption spectrum are composed of the Soret bands of Chl \textit{a} and \textit{b}, and broad, vibronically-structured absorption bands of several carotenoids  \citep{Liu_LHCII_PSII}, which approximately align with our predicted absorption pair at $\lambda_{0}=443$ nm. The red peaks arise from the $Q_{y}$ bands of Chl \textit{a} and \textit{b}, which are roughly in the same place as one of the weaker solutions in Fig. \ref{fig:Solar_peaks} \textbf{b.} We should note that the agreement is, at best qualitative and even then do not represent unique solutions. 

\subsection{Optimal antenna configurations for different stellar spectral fluxes}

We apply the $\Pi_{in}$ maximization procedure to stellar spectral fluxes in the $2300$ K $  \leq T_{s} \leq 5800$ K range, with $T_{s}=2300$ K representing the lower limit of the M-class stars. We consider incident wavelengths only in the $200$ nm $< \lambda_{0} < 1200$ nm range as this covers (with a wide margin) all Earth-based photo-autotrophs. We do not consider any atmospheric modulation of $I_{s}$, focusing instead on general qualitative trends. To reduce the parameter space (and because narrower $A(\lambda)$ performed better for the $5800$ K spectrum we consider only $\sigma = 10$ nm.   

\begin{figure*}
\centering
\includegraphics[width=0.6\linewidth]{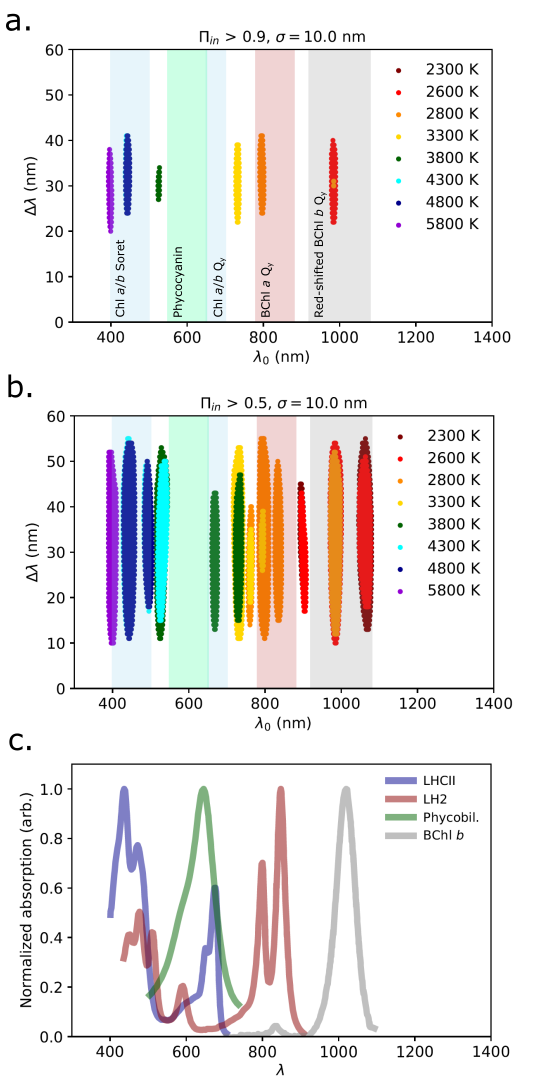}
\caption{\textbf{a.} All antenna configurations ($\lambda_{0},\Delta\lambda$) for which $\Pi_{in}\geq 0.9$ as a function of $T_{s}$. For comparison we show, as variously coloured bands, the approximate absorption regions of the Chl \textit{a} and \textit{b} Soret and $Q_{y}$ bands typical of higher plants, phycocyanin, one of the antenna pigments in the phycobilisome antenna of cyanobacteria, the BChl \textit{a} $Q_{y}$ band which captures light for purple bacteria, and the extremely red-shifted BChl \textit{b} $Q_{y}$ band from the antenna of some NIR-adapted purple bacteria. \textbf{b.} All antenna configurations ($\lambda_{0},\Delta\lambda$) for which $\Pi_{in}\geq 0.5$. \textbf{c.} Absorption spectra of LHCII (blue, digitized from \protect\cite{LHCII_spectrum}), phycocyanin (green, from \protect\cite{Yong_Phyco}), the peripheral antenna of purple bacterium \textit{Rhodobacter sphaeroides}, LH2 (maroon, from \protect\cite{Papagiannakis2022}), and the extremely red-shifted, BChl \textit{b}-enriched LH1 antenna from \textit{Blastochloris viridis} (grey, from \protect\cite{NamoonBChlb}).}\label{fig:temp_dependence}
\end{figure*}

Fig. \ref{fig:temp_dependence} \textbf{a.} shows all antenna configurations  for which $\Pi_{in}(\lambda_{0},\Delta\lambda) \geq 0.9$ as a function of $T_{s}$. These configurations represent highly optimized antennae that strongly satisfy both the requirments for noise cancellation and maximum power input ($\Delta^{op}\sim\Pi_{in}\sim0.95$). As a reference Fig. \ref{fig:temp_dependence} \textbf{a.} shows ranges of $\lambda_{0}$ that approximately correspond to the absorption peaks of known photosynthetic pigments and antenna structures. For $T_{s}=4800$ and $5800$ K (K- and G-type stars respectively) the optimum antennae roughly correspond to the blue Soret band of Chl \textit{a} and \textit{b}. The Soret band is composed of transitions from the Chl ground state to several high-lying singlet excited states. In LHCII the Chl Soret band captures blue photons (see Fig. \ref{fig:temp_dependence} \textbf{c.}, blue line) though a large proportion of this energy is lost as the resulting excitations rapidly ($< 1$ ps) and non-radiatively relax to the first singlet excited state which composes the $Q_{y}$ band. The Chl \textit{a} $Q_{y}$ band, at $\sim 680$ nm, is central to oxygenic photosynthesis as it is the precursor to the charge-separated states that initiation the photo-redox chemistry in the RCs.  

For $T_{s}=3800$ K, the lower end of the K-type stars, there is a narrow band of optimal antenna configurations at $\lambda_{0}\sim 540$ nm, within the region of pigments like phycocyanin (see Fig. \ref{fig:temp_dependence} \textbf{c.}, green line), the major light-harvesting pigment of the \textit{phycobilisome} antenna of cyanobacteria. Cyanonbacteria are also oxygenic and excitation of phycocyanins similarly rapidly relaxes to the the same Chl \textit{a} $Q_{y}$ states in the same RCs. It is only at $T_{s}=3300$ K that our model predicts an antenna configuration similar to the Chl \textit{a} and \textit{b} $Q_{y}$ bands.

For $T_{s}\leq 2800$ K, the lower end of the M-dwarf range, the optimal antenna configurations start to resemble the anoxygenic photoautotrophs. For $T_{s}=2800$ K the optimum is within range of the Bacteriochlorophyll \textit{a} (BChl \textit{a}) $Q_{y}$ band. For reference, Fig. \ref{fig:temp_dependence} \textbf{c.} shows the $800$ and $850$ nm bands from the peripheral LH2 antenna complex from the purple bacterium \textit{Rhodobacter sphaeroides} (maroon line), both of which originate from BChl \textit{a}. For $T_{s}\leq 2600$ K the optimum antenna absorbs at $\lambda_{0}\sim 1000$ nm which is comparable to the extremely red-shifted BChl \textit{b} found in the antennae of purple bacterial such as \textit{Blastochloris viridis} (see Fig. \ref{fig:temp_dependence} \textbf{c.}, grey line).

In Fig. \ref{fig:temp_dependence} \textbf{b.} we consider $\Pi_{in}\geq 0.5$, corresponding to antenna that are highly optimized to \textit{either} noise cancelling \textit{or} power input, or somewhat optimized for both ($\Delta^{op}\sim\Pi_{in}\sim 0.7$). We see the general temperature dependence is the same but with a broader variance. For example, for both $T_{s}=3800$ and $3300$ K there are a set of less optimal solutions in the Chl \textit{a} and \textit{b} $Q_{y}$ regions. For $T_{s}\leq 2600$ K there are less optimal solutions in the $\lambda_{0}\sim 900$ nm region, though on Earth no organisms harvest light in this region as these photons are strongly absorbed by atmospheric water \citep{stevens2013waterin}.  

\subsection{Comparing absolute power input for optimal antennae around different stars}

The estimated habitable zones for each $T_{s}$ are listed in Table \ref{tab:habitable}. We note that $\langle A_{p}\rangle(T_{s}=5800 \text{K}) \sim 0.75 \text{AU}$, reflecting our neglect of any atmospheric greenhouse effect in our calculation of radiative equilibrium. Still the estimates of $\langle A_{p}\rangle(T_{a})$ are qualitatively reasonable.  

\begin{table*}
\begin{tabular}{ | c c c c c |}
$T_{s}$ (K) & $R_{s}/R_{\odot}$  & $A_{p}^{max}$ (AU) & $A_{p}^{min}$ (AU) &	$\langle A_{p}\rangle$ (AU) \\
\hline
$2300$ & $0.117$ &	$0.019$ & $0.010$ & $0.015$ \\
$2600$ & $0.133$ & $0.028$ & $0.015$ & $0.021$ \\
$2800$ & $0.254$ & $0.062$ & $0.033$ &	$0.048$ \\
$3300$ & $0.299$ & $0.102$ & $0.055$ & $0.078$ \\
$3800$ & $0.613$ & $0.276$ & $0.148$ & $0.212$ \\
$4300$ & $0.694$ & $0.400$ & $0.214$ & $0.307$ \\
$4800$ & $0.775$ & $0.557$ & $0.299$ & $0.428$ \\
$5800$ & $0.936$ & $0.982$ & $0.526$ & $0.754$ \\
\hline   
\end{tabular}\caption{Table of estimated habitable orbital radii, $A_{p}$, as a a function of effective stellar temperature, $T_{s}$, and radius, $R_{s}$ (relative to the Solar radius, $R_{\odot}$). $A_{p}^{max}$ corresponds to an equilibrium planetary temperature of $T_{p}=273$ K, while $A_{p}^{min}$ corresponds to $T_{p}=373$ K. $\langle A_{p}\rangle$ is the mid-point between these distances.}\label{tab:habitable}
\end{table*}

\begin{figure*}
\centering
\includegraphics[width=0.6\linewidth]{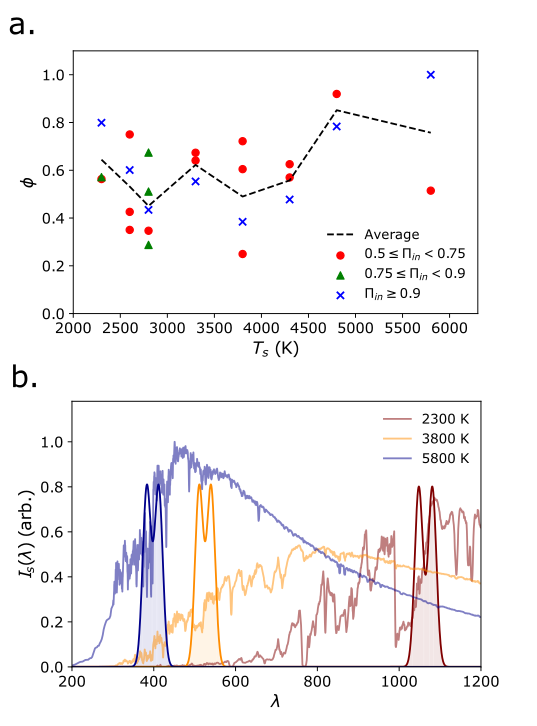}
\caption{\textbf{a.} The normalized input power, $\phi$, (as defined in eqn. \ref{eq:7}) for optimal antenna configurations as a function of $T_{s}$, assuming $\sigma=10$ nm in all cases. The reference value is the absolute optimum antenna configuration for $T_{s}=5800$ K ($\lambda_{0}=397.5$ nm, $\Delta\lambda =29.0$ nm). The dashedline indicates the average $\phi$ across all identified configurations for a given temperature. \textbf{b.} For illustrative purposes we show the optimum antenna absorption profiles, $A(\lambda;\lambda_{0},\Delta\lambda)$, for $T_{s}=2300,$ $3800$,and $5800$ K. For clarity the normalized $A(\lambda)$ have been resclaed by a factor of 40. In each case the requirement for a noise-cancelling antenna (\protect\cite{Arp2020}) selects configurations located on the slope of $I_{s}$}\label{fig:power_input}
\end{figure*}

Fig. \ref{fig:power_input} \textbf{a.} plots the relative input power, $\phi$, as defined in Eqn. \ref{eq:7} and with the optimum antenna configuration for $T_{s}=5800$ K (see Fig. \ref{fig:5800K_maps_peaks} \textbf{c.} as the reference configuration. Across the entire range of $T_{s}$ $\phi$ decreases by no more than a factor of $\sim 0.2$. This reflects the fact that the assumption of habitable planetary temperatures means that low stellar luminosity, 
\begin{equation}\label{eq:10}
L_{s}(R_{s},T_{s})=4\pi R_{s}^{2}\int_{0}^{\infty}d\lambda F_{s}(\lambda;R_{s},T_{s})
\end{equation}
is compensated for by small orbital radii, $A_{p}$. This implies that, in terms of the pure input power from a photosynthetic antenna, light limiting conditions are not necessarily a feature of exo-planets orbiting M-dwarf stars. 

\section{Discussion}

Since we are discussing the absorption characteristics of hypothetical photosynthetic organisms,and since we are interested in general trends rather than specific exo-planetary systems, we have adopted a relatively simple model.
We balance two considerations, the need to minimize input noise and the need to maximize overall energy input. In terms of the validity of the method tested this first on a $T_{s}=5800$ K model stellar spectral flux and then on a standard sea-level terrestrial Solar spectrum. In both cases our model predicts optimal antenna configurations qualitatively similar to the Chl \textit{a.} and \textit{b.} Soret bands. This is also true if, as \cite{Arp2020}, we consider only the requirement for noise cancellation. In neither case do we unambiguously predict anything resembling the Chl \textit{a.} and \textit{b.} bands. As \cite{Arp2020} acknowledged, for a realistic incident spectrum there are many local peaks in $\Delta^{op}$ and their reported fits to the antennae of higher plants are obtained by matching solutions from a selection of many. It should be noted that we employed a different method for identifying local maxima in $\Delta^{op}$, $P_{in}$ and $\Pi_{in}$, compared to \cite{Arp2020} (persistence homology rather than variations in curvature). Moreover, we explicitly compare the relative amplitude of these peaks rather than treating all local maxima as equivalent solutions. We found that $Q_{y}$-like antenna configurations were generally topologically indistinct and sub-optimal ($\Delta^{op}$ and $\Pi_{in} < 0.5$).    

The Chl Soret band functions as an \textit{accessory} light-harvesting pathway in photosynthetic antennae, capturing blue photons ($\lambda < 500$ nm) and converting them to usuable lower energy quanta ($\lambda\sim 700 $ nm) via non-radiative decay to the $Q_{y}$-band. \cite{Kiang2007} identified the Soret band as in their rules of the evolutionary relationship between antennae and spectral irradiance, that antennae have accessory pigments that absorb on the blue edge of the irradiance window. The Chl $Q_{y}$-bands are the \textit{primary} antenna states representing the $\lambda\sim 700$ nm quanta required by the RCs and the the immediate precursor states to the primary charge separation event that initiates the photosynthetic light-reaction. Clearly, $Q_{y}$-bands in the antenna are not (strongly) selected by the evolutionary pressures of noise cancellation and maximizing power input, but by the redox requirements of the RCs. These in turn may have been dictated by the availability of electron sources (such as water) and electron acceptors such as quinones and Fe/S clusters \citep{Kiang2007}. Alternatively, recent quantum chemical simulations suggests that 'primitive' tetrapyrole precursor molecules may have existed on Earth before life evolved. \textit{Phot0}, which resembles the metal-coordinating inner ring of Chl, has a Soret-$Q_{y}$ absorption profile similar to Chl and could have formed abiotically in the reducing conditions of early Earth \citep{Phot0}. Therefore, this absorption pattern could have been evolutionarily 'locked-in' long complex photosynthetic pathways.      

In line with \cite{LehmerFarRedEdge2021}, applying this model to decreasing $T_{s}$ reveals optimal antennae configurations at increasing wavelengths, tracking the blue edge of the spectral irradiances and it progressively red-shifts. At around $T_{s} = 3300$ K the optimum antenna reaches the energetic lower-limit of oxygenic photosynthesis ($\lambda_{0}\sim 700-750$ nm). In this region we find the absorption maxima of BChl \textit{c--f} pigments in the antennae of various green sulfur bacteria \citep{Gregersen_Green_Sulfur, Bryant2006Prokaryote, Bryant2007CandidatusCT, VoglChlf}. However, there are also species of oxygenic cyanobacteria that harvest light in $700-800$ nm region of the spectrum using  Chl \textit{d} and \textit{f}\citep{Tros2021RedLimit,MascoliRedLimit}. Although it is still a matter debate, there is compelling evidence that these far-red adapted cyanobacteria still possess the conventional $680$ (PSII) and $700$ nm (PSI) RCs, meaning that that $700-800$ nm photons absorbed by the antenna are transferred 'uphill' to the RCs against a thermodynamic barrier. These organisms therefore sacrifice antenna efficiency for the ability to collect far-red photons for which there is reduced competition with other oxygenic photoautotrophs \citep{Mascoli_FarRed_RC}.

At $T_{s}=2800$ K the optimal antenna overlaps with BChl\textit{a} and \textbf{b} from purple bacteria ($805 - 890$ nm). As an example, in Fig. \ref{fig:temp_dependence} \textbf{c.}, we show the absorption profile of the LH2 antenna complex from \textit{Rhodobacter sphaeroides} which has two well-separated Gaussian peaks at approximately $800$ and $850$ nm \citep{Papagiannakis2022}. Both peaks originate from BChl \textbf{a}, with the $850$ nm band being red-shifted due to a combination of pigment-protein and pigment-pigment interactions, illustrating the inherent \textit{tunability} of the optical properties of antenna pigments. 

For the lowest effective temperatures, $T_{s}\leq 2600$ K, the optimal antennae configurations are in the region $\lambda_{0}=1000-1100$ nm. There are examples of NIR-capturing antennae in our biosphere, with extremophiles such as \textit{Blastochloris viridis} \citep{NamoonBChlb} and \textit{Ectothiorhodospira halochloris} \citep{STEINER1985} which possess antenna enriched in extremely red-shifted BChl \textit{a} and \textit{b} which capture light down to $1020$ nm (see Fig. \ref{fig:temp_dependence}\textbf{c.}). However, it should be noted that they also absorb in the $800-870$ nm region and still possesses the usual purple bacterial $870$ nm reaction centre. As with the far-red adapted cyanobacteria the antenna is transferring energy against a thermodynamic gradient. In this case the barrier is a seemingly prohibitive $8k_{B}T$ which may suggest an extreme trade-off between antenna efficiency and the ability to capture a very under-utilized region of the spectrum, or that energy transfer occurs in a regime in which the steady state approximation does not apply. 

Assuming a fixed antenna size/cross-section, we find that the input power delivered by these optimal antenna varies surprisingly little with stellar temperature, $T_{s}$ (see Fig. \ref{fig:power_input}). This is due, in part, to that fact that our model selects for maximal input power, but mainly because the requirement for 'habitable' surface temperatures requires high surface irradiance. There is some variation in input power but this is no more than a factor of $0.2$. This is an extremely small difference when compared to the variation in photon fluxes that support photosynthesis on Earth. Canopy plants can be subject to fluxes as high as $2000 \phantom{x}\mu\textrm{mol photons}\phantom{x} m^{-2} s^{-1}$ ($\sim 1.2\times 10^{21} \phantom{x}\textrm{photons}\phantom{x} m^{-2} s^{-1}$) while for shade-adapted species it can be $\ll 100 \phantom{x}\mu\textrm{mol photons}\phantom{x} m^{-2} s^{-1}$. If we include the anoxygenic organisms then there are species of green sulfur bacteria that thrive in fluxes as low as $10^{-7} \phantom{x}\mu\textrm{mol photons}\phantom{x} m^{-2} s^{-1}$ from deep sea hydrothermal vents \cite{BeattyGSBhydrothermal}. This raises a very important feature of photosynthetic antenna that our model (and others) completely neglects: that antenna structure and cross-section are variable. 

While the photosynthetic RCs are generally highly-conserved, antenna structure is very diverse (see Fig. \ref{fig:temp_dependence}). Moreover, they are flexible, in part due to their modular and hierarchical structures. For example, higher plants (tracheophytes) all possess the same antenna structures, RC-coupled light-harvesting complexes embedded in the internal membrane structures (thylakoids) chloroplasts. Still different species evolve in very different light environments by altering leaf size and orientation, chloroplast density, thylakoid size, etc. \citep{Bateman_Physiology}. Flexibility exists within a single species, with plants like ivy (\textit{Hedera helix}) can acclimate to both high light ($\sim 1000 \phantom{x}\mu\textrm{mol photons}\phantom{x} m^{-2} s^{-1}$) and low light ($<100 \phantom{x}\mu\textrm{mol photons}\phantom{x} m^{-2} s^{-1}$), though growth rate is limited at the lower limit and photodamage is problem at the upper \citep{Oberhuber1991}. The ability of photosynthetic organisms to \textit{acclimate} to variable light environments is relevant to exo-biology. Recently, \cite{Battistuzzi_2023} demonstrated that two species of cyanobacteria (the model organism \textit{Synechocystis sp.}PCC 6803 and the far-red adaptable \textit{Chlorogloeopsis fritschi}) could acclimate to the spectral fluxes typical of low-mass ($T_{s}\sim 2800$ K) M-dwarf stars. They did this by up-regulating pigment synthesis, effectively increasing antenna cross-section, allowing them to maintain normal growth rates. While this is acclimation of an organism to low stress, and not the evolution of an organism specifically to suit these light-conditions, it does indicate that our model (and previous) may be too pessimistic in predicting the range of light-conditions for photosynthesis.     

Another problem with models based purely on average incident spectral flux is they neglect wider considerations of habitability. We would actually expect a large proportion of the habitable-zone planets orbiting M-dwarf stars to be tidally-locked \citep{KASTING1993Tidal,Barnes_Tidal}. Of these a significant fraction may therefore lack the tectonic activity needed to maintain a carbon cycle \citep{cockell2016habitability} which in turn would preclude the atmospheric stabilization needed to maintain liquid water \citep{McIntyre_2022}. While anoxygenic photoautotrophs don't exploit water as an electron source, they do require it a the universal intra-cellular solvent. Even if liquid water does exist it is not clear how the lack of a diurnal cycle would impact photosynthesis. On Earth, photosynthetic orgsanisms depend on a diurnal cicadian clock \citep{Dvornyk2016EvolutionOT} and there is some evidence that the evolution of oxygenic photosynthesis was influenced by increasing day length on young Earth \citep{Klatt_Diurnal}. Conversely, \cite{tang_vincent_2000}, in experiments on cyanobacteria, showed that net photosynthetic productivity decreases significantly in the absence of a period of darkness. The influence of tidal locking may therefore have as significant an impact on photosynthetic viability as incident light quality.      

In conclusion, though the question of photosynthetic viability around low-mass stars is extremely complex and multivariate, we have shown that lack of light is most likely not a reason to exclude it. However, around the lowest mass stars photosynthesis may resemble Earth's anoxygenic bacteria rather than complex oxygenic organisms. However, an immediate extension of this work will be a more careful consideration of antenna structure that factors n the inherent adaptibilty of Earth's photosynthetic light-harvesters.

\section*{Acknowledgments}

TJH is funded by a Royal Society Dorothy Hodgkin Fellowship, which also funded a summer studentship of GMMC. CDPD is funded by BBSRC (BB/T000023/1).

\section*{Data Availability}
The python scripts used to produce the models presented in this paper are available on request to Christopher Duffy (\href{mailto:c.duffy@qmul.ac.uk}{c.duffy@qmul.ac.uk})



\bibliographystyle{mnras}
\bibliography{references} 







\bsp	
\label{lastpage}
\end{document}